\documentclass{aa}
\usepackage[utf8]{inputenc}   
\usepackage[T2A,T1]{fontenc}  
\usepackage[english]{babel}   
\usepackage{aas_macros}
\usepackage{ulem}
\usepackage{hyperref}
\hypersetup{
    colorlinks=true, 
    linktoc=all,     
    linkcolor=blue,  
    urlcolor=magenta,
    citecolor=[RGB]{0,128,0},
}
\usepackage{graphicx}         
\usepackage{float}            
\usepackage{caption}          
\usepackage{afterpage}        
\usepackage{placeins}         
\usepackage{caption}          
\usepackage{pgf,tikz}
\usetikzlibrary{arrows}
\usepackage{multirow}
\usetikzlibrary{shapes}
\usetikzlibrary{snakes}
\usepackage{import}
\usepackage{amsmath}          
\usepackage{amssymb}          
\usepackage{nicefrac}         
\usepackage{gensymb}          
\usepackage{txfonts}

\begin{document} 

\title{Testing General Relativity with geodetic VLBI: what profit from a single, specially designed experiment?}
\titlerunning{General Relativity test with VLBI}
\authorrunning{Titov et al.}
\author{
O. Titov\inst{1} 
\and
A. Girdiuk\inst{2}
\and
S. B. Lambert\inst{3}
\and
J. Lovell\inst{4}
\and
J. McCallum\inst{4}
\and
S. Shabala\inst{4}
\and
L. McCallum\inst{4}
\and
D. Mayer\inst{2}
\and
M. Schartner\inst{2}
\and
A. de Witt\inst{5}
\and
F. Shu\inst{6}
\and
A. Melnikov\inst{7}
\and
D. Ivanov\inst{7}
\and
A. Mikhailov\inst{7}
\and
S. Yi\inst{8}
\and
B. Soja\inst{9}
\and
B. Xia\inst{6}
\and
T. Jiang\inst{6}
}

\institute{
Geoscience Australia, Canberra, PO Box 378, ACT, 2601, Australia, \email{oleg.titov@ga.gov.au}
\and
Department of Geodesy and Geoinformation, Research Group Advanced Geodesy, TU Wien, Gusshausstra{\ss}e 27-29/E120.4, Wien-1040, Austria
\and
SYRTE, Observatoire de Paris, Universit\'e PSL, CNRS, Sorbonne Universit\'e, LNE, Paris, France
\and
University of Tasmania, Private Bag 37, Hobart, Tasmania, 7001, Australia
\and
Hartebeesthoek Radio Astronomy Observatory, PO Box 443, Krugersdorp, 1740, South Africa
\and
Shanghai Astronomical Observatory, 80 Nandan Road, Shanghai, 200030, China
\and
Institute of Applied Astronomy, Kutuzov Embankment, 10, Saint-Petersburg, 191187, Russia
\and
National Geographic Information Institute, Space Geodetic Observatory, Sejong, PO Box 30060, South Korea
\and
Jet Propulsion Laboratory, California Institute of Technology, 4800 Oak Grove Drive, Pasadena, CA 91109, USA
}

\date{}

\abstract
{\bf We highlight the capabilities of the geodetic VLBI technique to test General relativity in the classical astrometric style, i.e., measuring the deflection of light in the vicinity of the Sun.}
{\bf In previous studies, the parameter $\gamma$ was estimated by global analyses of thousands of geodetic VLBI sessions. Here we estimate $\gamma$ from a single session where the Sun has approached two strong reference radio sources 0229+131 and 0235+164 at an elongation angle of 1-3$^{\circ}$.}
{\bf The AUA020 VLBI session of 1 May 2017 was designed to obtain more than 1000 group delays from the two radio sources. The Solar corona effect was effectively calibrated with the dual-frequency observations even at small elongation from the Sun.}
{\bf We obtained $\gamma$ with a precision better than what is obtained through global analyses of thousands of standard geodetic sessions over decades. Current results demonstrate that the modern VLBI technology is capable of establishing new limits on observational test of General Relativity.}
{}


\maketitle

\section{Introduction}

Very long baseline interferometry (VLBI) measures the difference in arrival times (known as group delay) of radio waves at two radio telescopes from distant radio sources with a precision of 20-40 ps \citep{2012JGeo...61...68S}. The observations are carried out at two frequency bands: 2.3~GHz (S-band) and 8.4~GHz (X-band); the lower frequency is used to calibrate ionospheric fluctuations in X-band data. By combining many years of observations, this technique is capable of producing very accurate positions of the reference radio sources. E.g., the error floor of the current realization of the fundamental celestial reference frame, the ICRF2, is 40~$\mu$as \citep{1538-3881-150-2-58}. For one standard single geodetic VLBI experiment, positions of radio sources are estimated with an accuracy of about 0.1 to 1~mas.

In accordance with General Relativity \citep{1916AnP...354..769E} the radio waves slow down due to the gravitational potential of the Sun \citep[the so-called Shapiro effect; see][]{Shapiro1964,1967Sci...157..806S}, making VLBI a useful tool for testing General Relativity by means of the parameterized post-Newtonian (PPN) formalism \citep{1993tegp.book.....W}. Nevertheless, the accuracy of the PPN parameter $\gamma$ obtained from absolute or differential VLBI observations \citep{2009ApJ...699.1395F,LambertLePoncin-Lafitte2009,LambertLePoncin-Lafitte2011} remains worse than the current best limit of $(2.1\pm2.3)\times10^{-5}$ based on Cassini radio science experiment \citep{2003Natur.425..374B} by an order of magnitude. The upper limits on the parameter $\gamma$ have been improved substantially in the past 30 years \citep{1984Natur.310..572R,1991Natur.349..768R,1995PhRvL..75.1439L,2003ApJ...598..704F}, but some authors \citep{2004PhRvL..92l1101S,LambertLePoncin-Lafitte2009} found degradation in the estimates of $\gamma$ with elongation, and suggested that this systematic effect may limit the improvement in the VLBI-derived $\gamma$ upper limits, despite the dramatic growth in the number of observations in recent decades.

The current paper focuses on radio source approaches at angular distances less than three degrees from the centre of the Sun in order to measure the light deflection effect at the highest magnitude and, thus, to avoid a possible bias caused by observations at larger elongations. We report on a special VLBI session, AUA020, that was run in May 2017 and on the single-session estimates of $\gamma$.

\section{Data}

The gravitational delay calculated as the difference between two Shapiro delays \citep{1990SvA....34....5K,1538-3881-125-3-1580,1991AJ....101.2306S,2010ITN....36....1P} is given by
 \begin{equation}
\tau_{\mathrm{grav}} = \frac{(1+\gamma)GM}{c^3} \ln \frac{\lvert \vec{r}_1 \rvert + \lvert \vec{r}_1 \cdot \vec{s} \rvert}{\lvert \vec{r}_2 \rvert + \lvert \vec{r}_2 \cdot \vec{s} \rvert},
\end{equation}
where $G$ is the constant of gravitation, $M$ is the mass of a deflecting body, $c$ is the speed of light in vacuum, $\vec{s}$ is the unit vector in the direction of the radio source, and $\vec{r}_1$ and $\vec{r}_2$ are geocentric distances of the two radio telescopes, $\lvert \vec{r}_1 \rvert$ means the vector $\vec{r}_1$ length and denotes the dot product. The gravitational delay is linked to the light deflection angle \citep{Titov_Girdiuk_2015} and is used to be observed with optical experiments \citep{DysonEddingtonDavidson1920}.

\begin{figure}
\centering
\includegraphics[width=\hsize]{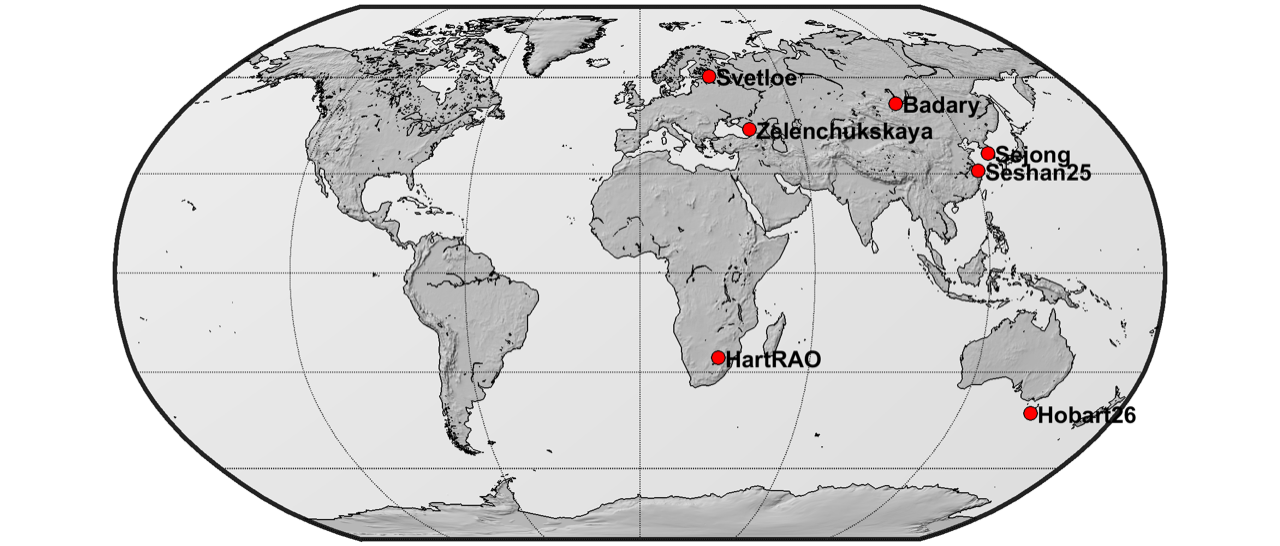}
\caption{Geometry of the AUA020 session network.}
\label{Fig_Network}
\end{figure}

\begin{figure*}
\begin{flushleft}
\includegraphics[width=\hsize]{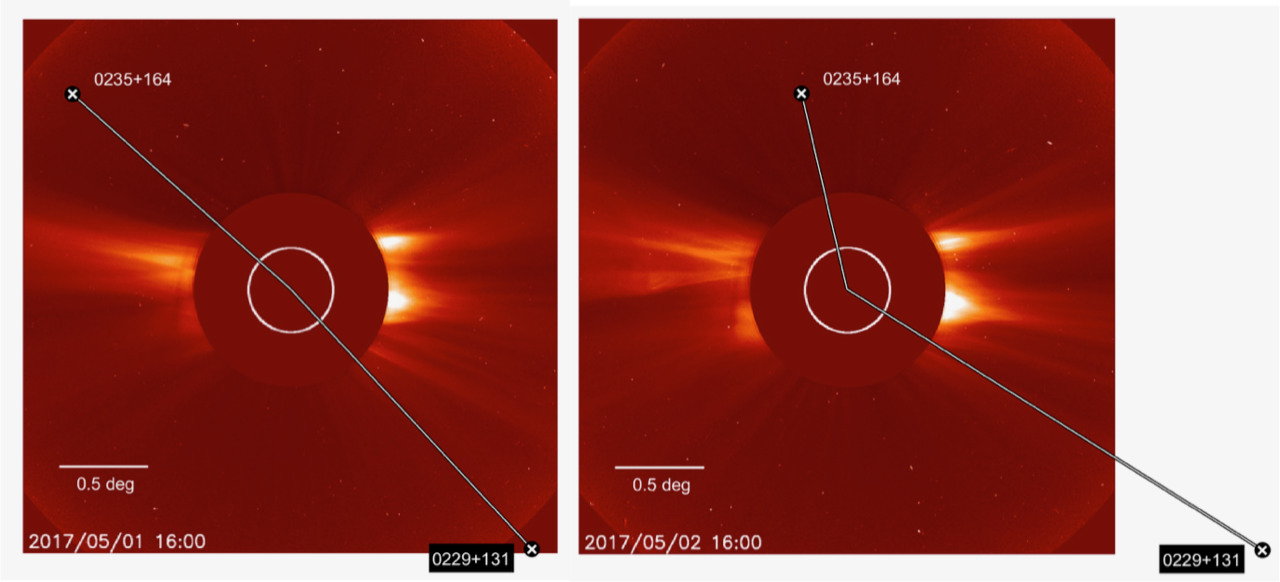}
\end{flushleft}
\caption{Geometry of the radio sources close to the sun at the start (Left) and at the end (Right) of VLBI session AUA020 with respect to a LASCO C2 image of the solar corona \citep{1995SoPh..162..357B}. The Sun is hidden behind the occultation disc of the coronagraph, with the white inner circle representing the limb of the Sun. The field-of-view is 1.5 degrees elongation.}
\label{Fig_Sun}
\end{figure*}

A dedicated geodetic VLBI experiment (AUA020, 01-02 May, 2017, part of AUSTRAL program) was scheduled to probe the gravitational delay effect using a network of seven radio telescopes (Svetloe, Zelenchukskaya, Badary, HartRAO, Seshan25, Sejong, and Hobart26) as shown in Fig.~\ref{Fig_Network}. Two radio sources 0229+131 and 0235+164 were observed at range of angular distances from 1.15$^\circ$ to 2.6$^\circ$ from the Sun. The position of both radio sources with respect to the Sun at the start of the experiment is shown on Fig.~\ref{Fig_Sun}. A serious issue in such a configuration is the solar thermal noise that penetrates to the signal through the side lobes, and could cause loss of data due to striking the signal-to-noise ratio. To overcome the problem, one has to

1) select strong radio sources with larger correlated flux density in both frequency bands,

2) use large radio telescopes with narrow side lobes and better sensitivity, and

3) use the highest possible data rate recording (e.g., 1 Gbps) to gain a better signal-to-noise ratio during the same integration time.

The schedule of AUA020 was designed to gain as many observations of the two encountered radio sources as possible. The previous attempts to observe reference radio sources close to the Sun used the standard scheduling strategy of geodetic experiments. In this mode all sources around the sky are observed a few times for 24 hours to provide a homogeneous sky coverage for each 1-hour time span. Therefore, the total number of delays of the sources within 2 degrees from the Sun was about 10-20 per session. The current schedule was customly prepared using the geodetic VLBI scheduling and analysis software VieVS \citep{BohmBohmBoisits2018}. From 16 to 22 UT, 1-May the standard geodetic mode was used. From 22 UT, 1-May to 15 UT, 2-May the schedule was focused on 0229+131 and 0235+164 repeating the following pattern:

1) scan of 0229+131,

2) scan of 0235+164,

3) two standard geodetic scans to improve the sky coverage,

4) startover again with 0229+131 and so on.

{\bf Lowest elevation was $5^{\circ}$.} During the final hour from 15 UT until 16 UT, 2-May the standard geodetic scheduling mode was applied again. The idea was that using this fixed pattern one will have a short slew time between 0229+131 and 0235+164 which increases the total number of scans. Using the strategy, 0235+164 was scheduled 108 times with 846 observations and 0229+131 was scheduled 109 times with 821 observations. Therefore, in spite of substantial loss of data, the total number of good delays exceeded 1000, making the session really outstanding among all others for testing of general relativity.

The target radio source 0229+131 is a defining source of the ICRF2 whose position is given with an accuracy close to the ICRF2 noise floor of 40~$\mu$as. The position of the second target 0235+164 is less accurate by a factor of five but still at the level of the ICRF2 median error and largely below the millisecond of arc. Both sources are compact and their structure indices measured at the time of the ICRF2 work were of 2.4 and 1.3, respectively \citep{MaAriasBianco2009}, ensuring a structure delay lower than 2~ps \citep{FeyCharlot1997}.

\section{Analysis}

For purpose of cross-checking the results and testing their robustness, we processed the VLBI session AUA020 within two independent teams with two independent geodetic VLBI analysis software packages. The duplication of the analyses with two software packages also allows to use some specific options that are available on only one of them. The first analysis package is OCCAM \citep{2004ivsg.conf..267T} that implements the least-squares collocation method \citep{2000ITN_Titov} for calibrating the wet troposphere fluctuations, and to account for the mutual correlations between observables. The second one is Calc/Solve \citep{MaClarkRyan1986}, developed and maintained by the geodetic VLBI group at NASA GSFC, that uses classical least-squares. The modeling of intraday variations of the troposphere wet delay, clocks, and troposphere gradients is realized through continuous piecewise linear functions whose coefficients are estimated every 10 mn, 30 mn, and 6 hours, respectively. The rest of the parameterization is identical for both analyses. A priori zenith delays were determined from local pressure values, which were then mapped to the elevation of the observation using the Vienna mapping function \citep{VMF1}. Station positions and velocities were fixed to ITRF2014 values \citep{JGRB:JGRB51713} and corrected from pressure loading effects using relevant loading quantities deduced from surface pressure grids from the U.S. NCEP/NCAR reanalysis project atmospheric global circulation model \citep{NCEP_NCAR} and from the FES2004 ocean tide model \citep{FES2004}. A priori Earth orientation parameters were taken from the IERS EOP 14 C 04 data and the IAU 2000A/IAU 2006 nutation and precession models \citep{MathewsHerringBuffett2002,CapitaineWallaceChapront2003}. Since VLBI measurements of the nutation leaves residuals with respect to IAU 2000A of about 0.2~mas in rms due to various mismodeled and unmodeled nutation components \citep{DehantFeissel-VernierVironEtAl2003}, we estimate offsets to the celestial pole coordinates.

{\bf Radio source positions were fixed to the ICRF2 coordinates \citep{1538-3881-150-2-58}. This option, that constitutes one important difference between our analysis strategy and, e.g., \cite{LambertLePoncin-Lafitte2009} was used because source coordinates are unlikely to vary during the 24 hours of the session, in constrast to the 0.1-mas scale displacements of the radio center of many radio sources that are observed in long-term VLBI analyses and associated with intrinsic phenomena \citep{FeyEubanksKingham1997,MacMillanMa2007} or the Galactic aberration \citep{Kovalevsky2003}, that could impact the adjustment of $\gamma$.}

{\bf Though all standard geodetic parameters were estimated using all scans, the parameter $\gamma$, whose adjusted values are reported in Table~\ref{tabsol}, was estimated alternatively using all scans and using only scans relevant to either 0229+131 (within 2-3$^{\circ}$ to the Sun), 0235+164 (within 1-2$^{\circ}$ to the Sun), or both sources.}

\begin{table}
\caption{Estimates of $\gamma-1$ for the session AUA020, in unit of 10$^{-4}$, along with the session $\chi^2$ and the postfit rms delay $r$ in ps.}
\label{tabsol}
\centering
\begin{tabular}{lrrrrr}        
\hline
\hline
\noalign{\smallskip}
 & & $\gamma-1$ & $\sigma_{\gamma}$& $\chi^2$ & $r$ \\
 & & $10^{-4}$ & $10^{-4}$& & \\
\noalign{\smallskip}
\hline
\noalign{\smallskip}
\multirow{10}{*}{OCCAM} & \multicolumn{5}{l}{All stations} \\ \cline{2-6}
\noalign{\smallskip}
& All scans        &   0.56 & 1.15 & 0.34 & 28 \\
& 0235+164          &   1.34 & 1.58 & 0.34 & 28 \\
& 0229+131          &  -1.54 & 3.41 & 0.34 & 28 \\
& Both                   &   0.53 & 1.14 & 0.34 & 28  \\
\noalign{\smallskip}
& \multicolumn{5}{l}{With Sejong downweighted} \\ \cline{2-6}
\noalign{\smallskip}
& All scans           &   0.91 & 0.94  & 0.27 & 21 \\
& 0235+164            &   1.64 & 1.29  & 0.27 & 21 \\
& 0229+131            &   0.32 & 2.83  & 0.27 & 21 \\      
& Both                     &   0.89 & 0.94  & 0.27 & 21 \\
\noalign{\smallskip}
\hline
\noalign{\smallskip}
\multirow{4}{*}{Calc/Solve}
 & All sources                         &  -0.22 & 1.10 & 0.84 & 26 \\
 & 0235+164                          &   1.85 & 1.48 & 0.84 & 26 \\
 & 0229+131                          &  -6.84 & 2.53 & 0.84 & 26 \\      
 & Both                                   &  -0.26 & 1.09 & 0.84 & 26 \\
\noalign{\smallskip}
\hline
\end{tabular}
\end{table}

\section{Results}

{\bf Uncertainties on $\gamma$ reported in Table~\ref{tabsol} lie between $1\times10^{-4}$ and $4\times10^{-4}$. Our estimates appear therefore as precise as that obtained from global solutions using thousands of VLBI experiments \citep{LambertLePoncin-Lafitte2009,LambertLePoncin-Lafitte2011}. The formal error is about two times lower when $\gamma$ is fitted to the observations of the radio source that is two times closer to the centre of the Sun (0235+164) than to that of its counterpart (0229+131). Using all scans returns a result similar to using only scans relevant to 0229+131 and 0235+164, confirm that only sources a low elongation can efficiently constrain the PPN parameter.} Solutions from both software packages are consistent within the standard errors. The difference of postfit rms between OCCAM and Calc/Solve might find its origin in the different modeling of the nuisance parameters (stochastic versus CPWL function).  No large systematics are detected except a $2.7\sigma$ deviation in the Calc/Solve solution when only 0229+131 is used and whose origin is unclear: as both solutions started from the same a priori, the issue could rather be in the estimation method or in the handling of troposphere/clock parameters.

It appears that during session AUA020, data in three channels at Sejong station were lost due to technical reasons. Therefore, we reprocessed the previous analyses after downweighting (but not suppressing) Sejong data.  {\bf (We could not test this option with OCCAM only since Calc/Solve does not handle downweighting.)} The postfit rms of the solution is significantly lowered. The formal error on $\gamma$ is marginally lowered down to $9\times10^{-5}$.

\begin{figure}[htbp]
\centering
\includegraphics[width=\hsize]{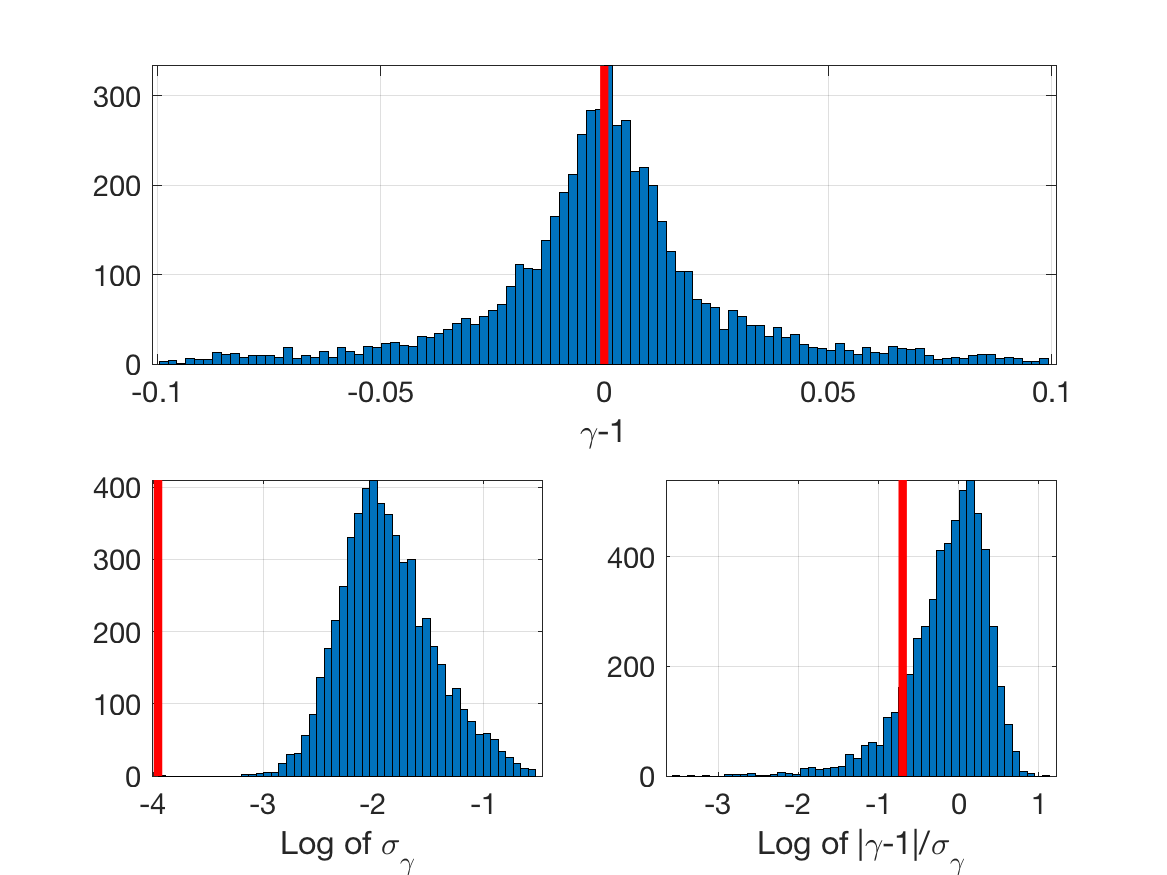}
\caption{Distributions of (Top) estimates of $\gamma-1$, (Bottom-left) their formal errors, and (Bottom-right) normalized estimates of $\gamma-1$ for all of the geodetic VLBI sessions. The vertical, red bar stands for the results of the AUA020 session.}
\label{figfulldb}
\end{figure}

For purpose of comparison of the AUA020 session with other standard geodetic VLBI sessions, we estimated $\gamma$ with Calc/Solve using the parameterization described above for each of sessions of the full geodetic VLBI data base made available by the International VLBI Service for geodesy and astrometry \citep[IVS,][]{NothnagelArtzBehrend2017} since 1979. The session list includes both single- and multi-baseline networks (at the exclusion of intensive sessions). The median postfit rms is 27 ps that is close to the postfit rms of the AUA020 session. The distribution of the obtained values of $\gamma-1$ is shown in Fig.~\ref{figfulldb} along with distributions of errors and normalized estimates. The distribution of errors in log-scale is slightly asymmetric, exhibiting a `tail' on its right side that might traduce results from sessions not designed for precise astrometry. Nevertheless, assuming a Gaussian shape, the log-scaled distribution peaks at 10$^{-2}$ with a $\sigma$ of $\sim0.5$. This makes the error estimate from AUA020, that is two orders of magnitude less, somewhat `outstanding'. The bottom-right panel of Fig.~\ref{figfulldb} shows that the major part of the sessions does not bring severe systematics, the estimates of $\gamma$ being unity within the error bars; session AUA020 is part of the session group that presents the lowest systematics.

We also processed solutions parameterized as in \cite{LambertLePoncin-Lafitte2009,LambertLePoncin-Lafitte2011}, thus estimating $\gamma$ as a global parameter over the same 6301 sessions, totaling 12.6 millions of ionosphere-free group delays. A priori positions for radio sources were taken from the ICRF2 \citep{1538-3881-150-2-58} and a no-net rotation constraint was applied to the defining sources. The postfit rms delay of the solution is 26 ps and its $\chi^2$ per degree of freedom is 1.00. We obtained $\gamma-1=(2.72\pm0.92)\times10^{-4}$ that yields a slight improvement with respect to \cite{LambertLePoncin-Lafitte2011} mainly due of the $\sim$5.3 millions of observations accumulated in the mean time. Removal of the session AUA020 led to $\gamma-1=(2.57\pm0.97)\times10^{-4}$, showing that AUA020 marginally - but still at a detectable level - improves the formal error at the level of $5\times10^{-5}$. However, these global solutions exhibit systematics at the level of 2-3$\sigma$ that may have their origin in spurious or unmodeled deformations of the celestial reference frame arising from, e.g., declination-dependent errors associated with the global network north-south asymmetry or the influence of troposphere gradient modeling \citep{MayerBoehmKrasna2017}, or other effects not yet addressed (e.g., Galactic aberration).

\section{Discussion and conclusion}

Although we detected no systematics on $\gamma$ within $10^{-4}$, several potential sources of error have to be considered.

{\bf The effect of the plasma of the solar corona on the VLBI observations was corrected to first order by a linear combination of the group delays in S and X bands. In order to investigate the magnitude of higher order dispersive effects, higher order terms were computed according to \cite{BassiriHajj1993}. The coronal electron density $N_e$ was parameterized by a radial power law, depending on the electron density $N_0$ at the Sun's surface, the radial falloff parameter $\beta$, and the distance from the Sun's center $r$ in solar radii: $N_e(r) = N_0r^{-\beta}$. We used the values $N_0 = 0.57\times10^{12}$ m$^{-3}$ and $\beta = 2$, which were estimated from dual-frequency VLBI observations during 2011 and 2012 \citep{SojaHeinkelmannSchuh2014}. Since the solar conditions during 2011 and 2012 were closer to a solar maximum \citep{SojaHeinkelmannSchuh2016} than during AUA020 on May 1, 2017, the selected model can be considered a worst case scenario. Additionally, a dipole model of the background solar magnetic field was assumed with the field strength of $10^{-3}$ Teslas at the poles. Using these models for the electron density and magnetic field, both terms of second and third order (proportional to the inverse third and fourth power of frequency, respectively) turned out to be well below 0.1~ps for the differential observations of VLBI, and thus negligible. A more sophisticated calibration could possibly be achieved by using more rigorous electron density and magnetic field models, taking into account regional differences. However, due to the small magnitude of the higher order terms, it is unlikely that such an effort would change the results concerning the relativistic investigations presented in this study.

Furthermore, we investigated the effect of ray path separation on the dispersive delay corrections. Following \cite{TylerBrenkleKomarek2030} (in particular, their Fig. 4), the different refractivity in S and X band causes a ray path separation of about 150 km at an elongation of 1.2$^{\circ}$ (4.5 solar radii). Assuming a projected baseline in radial direction from the Sun of 6000 km, the difference in ray separation is about 1 km between the rays of the two stations. Using the same model for the solar corona electron density as mentioned above, we compute a dispersive X band group delay of 7.05 mm for the ray path closer to the Sun and 6.97 mm for the ray path 6000 km further out. The differential delay is thus below 0.1 mm for this scenario, and should be even smaller for the actual observations during AUA020.

While the systematic effects based on radial or dipole models of the corona appear to be negligible, individual group delay observations are affected by random scatter caused by small-scale coronal structure and temporal variations thereof. Since these perturbations do not systematically affect the observations, we assume that they cancel out over the period of observations (17 hours with observations angularly close to the Sun). Since the ray paths to the radio sources 0235+164 and 0229+131 within small solar elongation happened to be in quiet regions (cf. Fig.~\ref{Fig_Sun}), the scatter was small enough that precise group delays could be successfully determined at such small elongations.}

The major source of stochastic noise in VLBI measurements resides in the unknown wet troposphere delay. The difference between VLBI estimates of the wet troposphere delay and independent radiometer data appears to stay within 3 mm, or 10 ps \citep{2013evga.conf..151T} suggesting that the impact of the wet troposphere delay on the astrometric light deflection angle estimation near the Sun is negligible.

The minimum angle between the Solar limb and the radio source capable of being observed is the most serious issue. We demonstrated here that the elongation of 1$^{\circ}$ is favorable for successful observations even for a radio source of moderate flux. Obviously, a further reduction of the elongation requires a trade-off between the higher flux of a potential candidate to be observed and exponential growth of the thermal noise reaching the receivers. The flux should be strong enough to pass through the medium surrounding the Sun without dissipation. Since there is no way to make an absolutely reliable prediction about the activity of the Solar corona in a particular area near the Solar limb, one has to try to schedule an experiment using the best possible initial conditions. Taking the best candidate for this kind of experiment, the best known radio source is 3C~279 (J1256-0574) that combines a small ecliptic latitude of 0.2$^{\circ}$ and flux density that is ten times stronger than for 0229+131 and 0235+164. As the signal-to-noise ratio (SNR) grows proportionally to the correlated flux density, the exponential increase of the thermal noise is likely to be compensated. Therefore, it appears reasonable to track the radio source 3C~279 in the range from 0.5$^{\circ}$ to 1.0$^{\circ}$ from the centre of the Sun. A successful detection of the signal at 0.5$^{\circ}$ would immediately result in an mprovement of uncertainties on $\gamma$ by a factor of two. Another experiment including 15 radio telescopes to observe the radio source 3C~279 at the same high data recording rate will collect at least 5000 single observations near the Sun (as the number of baselines increases from 21 to 105) with better precision of each single observation than during the May, 2017 experiment. Overall, a total improvement of the uncertainty on $\gamma$ by a factor of ten is expected, enabling to challenge the current limit imposed by the Cassini radio science experiment of \cite{2003Natur.425..374B}{\bf, although the Gaia astrometry on Solar system objects is expected to deliver an accuracy of $10^{-6}$ \citep{MignardKlioner2009,HeesLe-Poncin-LafitteHestroffer2017}.}

\begin{acknowledgements}
This paper is published with the permission of the CEO, Geoscience Australia. B. Soja's research was supported by an appointment to the NASA Postdoctoral Program, administered by Universities Space Research Association, at the Jet Propulsion Laboratory, California Institute of Technology, under a contract with National Aeronautics and Space Administration. We are grateful to D. Gordon (GSFC) for post-processing reduction of the AUA020 data.
\end{acknowledgements}

\bibliographystyle{aa} 
\bibliography{PPNpaperbib} 

\end{document}